\documentclass{jps-cp}
\usepackage{txfonts} 
\usepackage{comment}
\usepackage{color}
\usepackage{url}

\title{
Development of the Beam Monitor Detectors for the Low-E Beamline at the CERN SPS H2 Line
}

\author{
Y. \textsc{Hino}$^{1, {\dag}}$,
M. \textsc{Harada}$^{1, {\ddag}}$,
N. \textsc{Meguro}$^{1}$,
T. \textsc{Tada}$^{1}$,
Y. \textsc{Koshio}$^{1}$,
K. \textsc{Sakashita}$^{2}$,
M. \textsc{Friend}$^{2}$, \\
T. \textsc{Nakadaira}$^{2}$,
S. \textsc{Nishimori}$^{2}$,
and Y. \textsc{Nagai}$^{3}$
}

\inst{
$^{1}$ Okayama University, 3-1-1 Tsushima-naka, Kita-ku, Okayama, 700-8530, Japan \\
$^{2}$ High Energy Accelerator Research Organization (KEK), Tsukuba, Ibaraki 305-0801, Japan \\
$^{3}$Department of Atomic Physics, Et\"{o}v\"{o}s Lor\'{a}nd University, Budapest, Hungary \\
\quad \\
$^{{\dag}}$ now at KEK, Tsukuba, Ibaraki 305-0801, Japan \\
$^{{\ddag}}$ now at Kamioka Observatory, Institute for Cosmic Ray Research, University of Tokyo, Kamioka, Gifu 506-1205, Japan \\
}

\email{hino@post.kek.jp}

\recdate{\today}

\abst{
A series of beam tests were conducted at the KEK AR test beamline in order to investigate the performance of a prototype detector for instrumentation in the low-E beamline at the CERN H2 line.
For the silicon strip detector, the simultaneous readout of all strips of the X-Y detector was evaluated, in addition to the detection efficiency being assessed through coincidences with the trigger scintillators. 
The results demonstrated the successful acquisition of data with full strip readout and the reconstruction of the two-dimensional beam profile. 
Furthermore, we are contemplating the incorporation of GAGG crystals, an inorganic scintillator characterized by a high light yield ($\sim$$5 \times 10^{4}$~photons/MeV) and a brief time constant ($\sim$60~ns), within the newly developed time-of-flight detector.
We assessed the resolution of time-of-flight based on the data, with the objective of achieving the required performance benchmarks through future improvements to the time-of-flight detector.
}

\kword{Silicon strip detector, Scintillator, Neutrino Flux}

\begin{document}
\maketitle

\section{Introduction}
An understanding of the interaction between hadrons and materials at energies of $O(\mathrm{GeV})$ is crucial for various aspects of neutrino physics.
The T2K experiment, a long-baseline neutrino oscillation experiment, utilizes the decay-in-flight of pions and kaons as a result of the interaction between carbon and a 30-GeV proton beam to produce a neutrino beam.
The T2K experiment has made significant strides in reducing the uncertainty surrounding the flux of neutrinos by leveraging the findings of the NA61/SHINE experiment~\cite{cite:t2kflux}.
The NA61/SHINE experiment is a fixed-target experiment situated at the H2 line, employing beams extracted from the CERN SPS accelerator.
A complex of the time-projection chambers and bending magnets enables us to measure the momenta and kinds of produced particles, resulting in a differential cross section of hadron production crucial in the neutrino flux prediction~\cite{cite:na61}.

The H2 line has been demonstrated to reliably deliver hadron beams with a minimum energy of 13 GeV/c.
On the other hand, the extension of the lower limit of the beam momentum from 13 to 2 GeV/c will facilitate more precise measurements of neutrino oscillation~\cite{cite:spsc}.
These include the prediction of the atmospheric neutrino flux and the enhancement of the prediction of the flux of neutrinos in long-baseline oscillation experiments, such as T2K and Hyper Kamiokande (HK).
The measurement of $p + \mathrm{N} \to \pi^{\pm} + \mathrm{X}$ in the 2 to 13 GeV/c proton momentum range provides complementary phase-space data with the HARP high-angle datasets. This, in turn, leads to an improvement in the flux uncertainty from 8\% to less than 3\% for both muon and electron (anti-)neutrinos. 
Interactions between mesons of neutrinos and beamline materials, known as "meson rescattering," represent a significant source of uncertainty in the calculation of the flux of neutrinos in the sub-GeV range.
The low-energy beamline offers a valuable opportunity to measure pion interactions on aluminum, carbon, and iron at 3 and 8 GeV/c, thereby reducing the associated uncertainty to $\sim$1/5 of its original value.
This will result in the optimization of the sensitivity to the discovery of neutrino CP violation in the HK project, achieved by reducing the uncertainty of the ratio of the neutrino to anti-neutrino cross sections to less than 3\%.

The design of the low-E beamline is in progress and under review.
In addition to the construction of the beamline facility, the development of new detectors capable of monitoring beam position and particle identification is imperative.
This document presents the status of detector development, specifically focusing on a beam position detector and a Time-of-Flight (ToF) detector. 
These detectors are being evaluated based on the results of tests conducted using a 3~GeV/c electron test beam.

\section{Beam monitors for the low energy beamline}
The NA61/SHINE Japan group is responsible for the development of two detectors: a silicon strip detector (SSD) to measure the beam position and a scintillator-based time-of-flight (ToF) detector for particle identification.
In this section, specifications and requirements for the monitors will be described.

\subsection{Beam profile detector}
\begin{figure}[tb]
    \centering
    \includegraphics[width=0.6\textwidth]{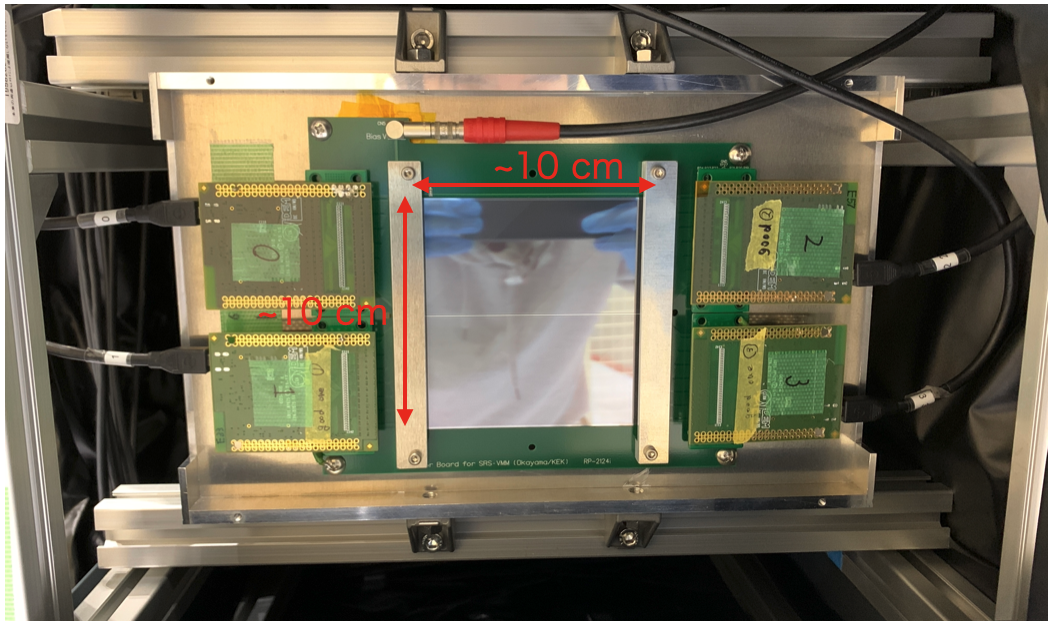}
    \caption{A photo of the X-SSD consisting of a Hamamatsu S13804 sensor and 4 APV hybrids. A source meter is used to supply bias voltage to the sensor and monitor the leak current simultaneously. The Y-SSD is deployed behind the X-SSD. \label{fig:ssd_photo}}
\end{figure}

In order to determine the momentum of the beam for each event, a series of beam profile detectors will be installed along the beamline at various locations. This method is analogous to the approach employed in other VLE beamlines at CERN~\cite{cite:h4vle}.
Additionally, a pair of SSDs will be positioned outside the vacuum downstream of quadrupole magnets, just before the target.
Thus, a prototype detector based on a Hamamatsu S13804 silicon strip sensor designed for the J-PARC g-2/EDM experiment has been developed (Fig.~\ref{fig:ssd_photo}).
The dimensions of the sensor are 98.77 mm $\times$ 98.77 mm, with a thickness of 320 $\mu$m. This design adequately covers the expected beam spot size of approximately 2 cm in both planes.
The integration of 512 strips, with a pitch of 190 $\mu$m, enables the attainment of enhanced position and angle resolution.
It is noteworthy that the realization of a 10 cm long strip has been accomplished through the implementation of the wire-bonding technique, whereby two strips have been bonded at readout pads that intersect at the center of the sensor, resulting in a 500 $\mu$m gap.
This design configuration gives rise to an insensitive line that traverses the center of each SSD.
Each strip sensor is connected to a readout pad on the board via AC coupling.

A signal from each strip is amplified and digitized by an APV hybrid module~\cite{cite:apv25} whose preamplifier gain is 64~mV/fC.
The expected charge of the signal due to a minimum ionizing particle (MIP) is $\sim$3.8~fC, according to the measurement~\cite{cite:g-2ssd}.
We employed 4 APV hybrid modules on a single SSD because an APV hybrid has 64 channels.
In reality, a pair of a vertically separated SSD (X-SSD) and a horizontally separated one (Y-SSD) is used to measure the two-dimensional position of a charged particle as it passes through.
The detector will be operated with 100~V of bias voltage, and the sensor will reach at full depletion, according to the specification document from Hamamatsu.

\subsection{Time-of-flight detector}
In order to achieve particle identification (PID) in the momentum range of 2-13 GeV/c, it is planned to utilize two different complementary PID detectors, such as threshold Cerenkov detectors and a ToF detector.
The minimum requirement of ToF resolution is 500~ps to achieve the required PID capability, separating proton, kaon, pion, and positron in case a couple of the threshold Cerenkov detectors are employed~\cite{cite:carlo, cite:carlo_phd}.
In the low-energy beamline, a reduction in the amount of material at the ToF detector is desired to minimize beam loss.

Two candidates for scintillation material in the ToF detector are currently under investigation.
One of these materials is a plastic scintillator, which is a well-established component in ToF measurement due to its rapid response.
The second candidate is a GAGG(Ce), cerium-doped gadolinium aluminum gallium garnet crystal, a scintillator material that is both non-hygroscopic and non-self-radioactive.
GAGG has a shorter time constant and higher light yield than other crystal scintillators, such as NaI(Tl), as illustrated in Table~\ref{tab:scint}. 
Therefore, it possesses the capacity for excellent time resolution due to the elevated photo statistics at the initial segment of the waveform.
A prototype ToF detector comprising a 0.5~mm-thick GAGG(Ce) crystal, integrated with an acrylic light guide and a 2-inch photomultiplier tube (PMT), was engineered to assess the viability of the GAGG-based ToF detector.

\begin{table}[t]
    \centering
    \caption{Property of candidate scintillators for the ToF detector as well as NaI(Tl) for reference. \label{tab:scint}}
    \begin{tabular}{ccccc}\hline
        Scintillator & Density /gcm${}^{-3}$ & Radiation length /cm & Time constant /ns & Light yield photons/keV \\\hline
        Plastic  & 0.9 & 50  & 3   & 10  \\
        GAGG(Ce) & 6.6 & 1.6 & 80  & 50  \\
        NaI(Tl)  & 3.7 & 2.6 & 230 & 40  \\\hline
    \end{tabular}
\end{table}

\section{Beam test at KEK AR test beamline}
The tests of the developed prototype detectors were performed using 3~GeV/c electron test beam provided at the AR test beamline (ARTBL) in KEK (Tsukuba) in March and December of 2023.
During the designated machine period, the beam was consistently supplied, and the event rate attained a level of over 1~kHz at the beam entrance within the experimental area.

\begin{figure}[tb]
    \centering
    \includegraphics[width=0.6\textwidth]{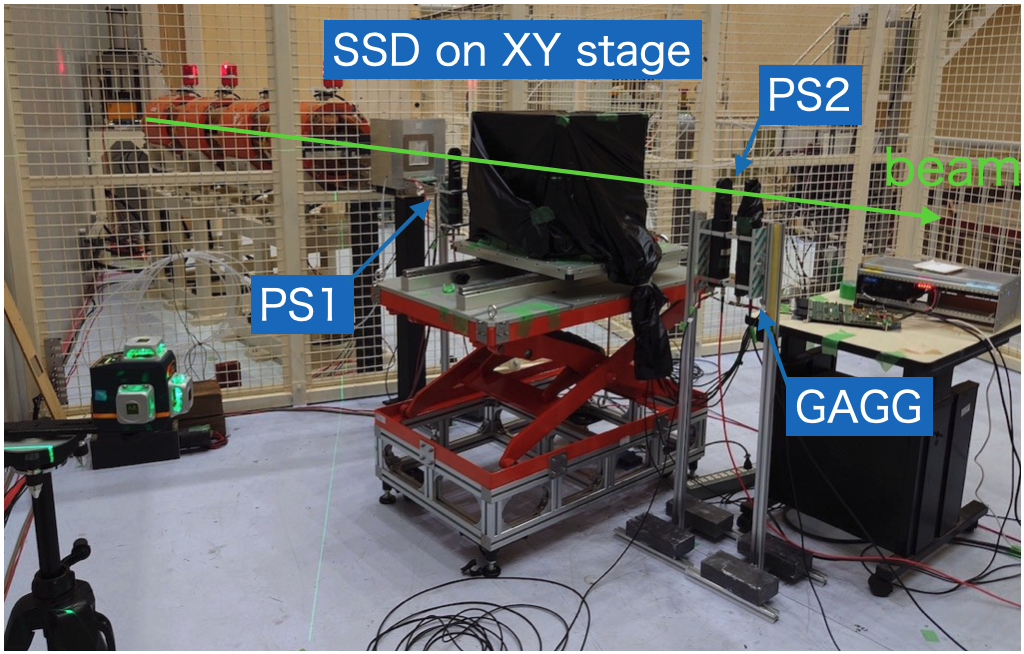} \\
    \includegraphics[width=0.9\textwidth]{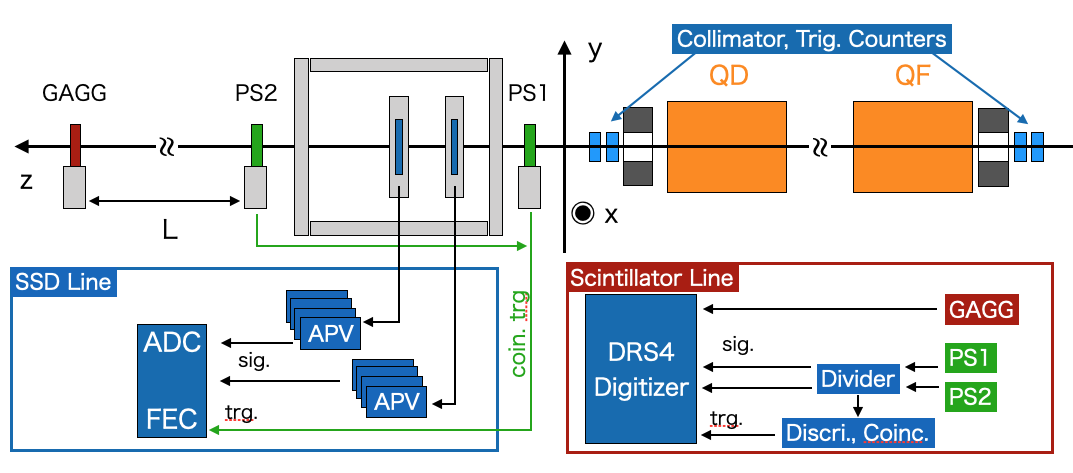}
    \caption{(Top) A photo of the experimental setup in the beam test at the KEK ARTBL. The electron beam was extracted from the beam shutter and was transported to the experimental area after the focus and defocus quadrupole magnets. (Bottom) The schematic illustration of the experimental setup. Both data of the SSDs and the scintillators were acquired by the trigger issued in coincidence with PS1 and PS2. \label{fig:setup_photo}}
\end{figure}

Figure~\ref{fig:setup_photo} displays an entire view of the experimental setup for the beam test (top) and its schematic illustration (bottom).
The SSD was placed in a dark box and positioned on the X-Y stage, which allows for vertical and horizontal adjustment of the beam spot on the silicon sensor with an accuracy of 0.1~mm.
Two detectors, each comprising a 1-cm-thick plastic scintillator coupled with an acrylic light guide and a 2-inch PMT (PS1, PS2), were positioned in front of and behind the SSD, respectively.
This configuration enables the assessment of the detection efficiency of the SSD with respect to MIP activity, contingent upon the triple coincidence of these detectors.
The prototype time-of-flight (ToF) detector, designated as GAGG, was positioned posterior to the PS2.
ToF measurements were subsequently executed, with distances between the PS2 and GAGG ranging from 0.1 to 2.3~m.
The alignment of the detectors was adjusted using a laser alignment tool based on the reference lines of the beamline. Consequently, the resulting uncertainty due to setup alignment was found to be negligible.

The signals from the SSDs were processed and recorded using a Scalable-Readout-System (SRS). 
The trigger for data taking was issued in coincidence with MIP activity in both PS1 and PS2.
The waveforms of GAGG, PS1, and PS2 were acquired by the DRS4 evaluation board v5~\cite{cite:drs4}, a waveform digitizer capable of sampling the four input signals simultaneously from 0.7 to 5~GHz with 1024 sampling points. 
In this beam test, a 1~GHz sampling rate was configured to obtain a 1024~ns window for each waveform.
The coincidence trigger was supplied to the DRS4 board for data taking.

\subsection{Q-V response of the silicon sensors}
\begin{figure}[tb]
    \centering
    \includegraphics[width=0.9\textwidth]{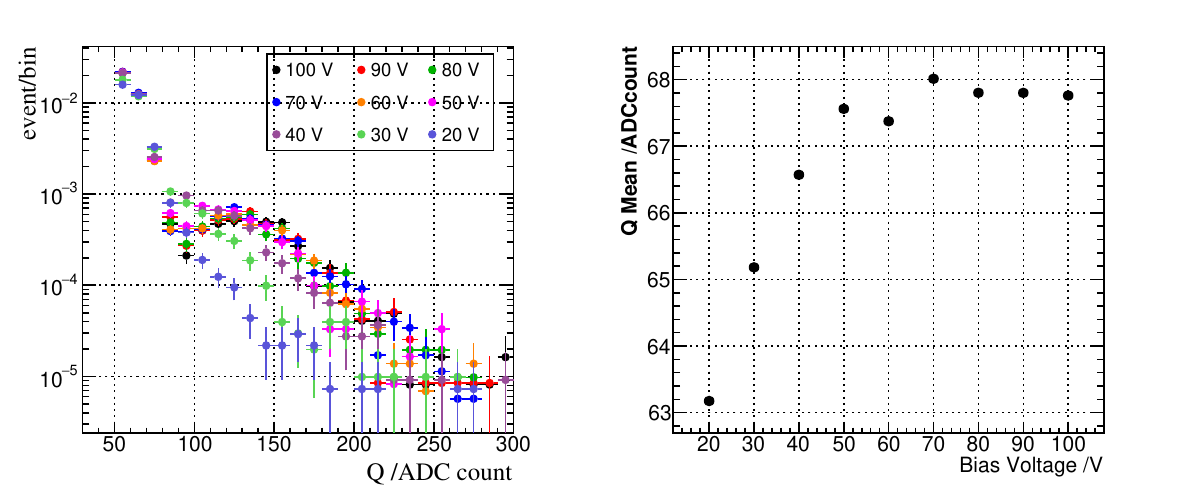}
    \includegraphics[width=0.9\textwidth]{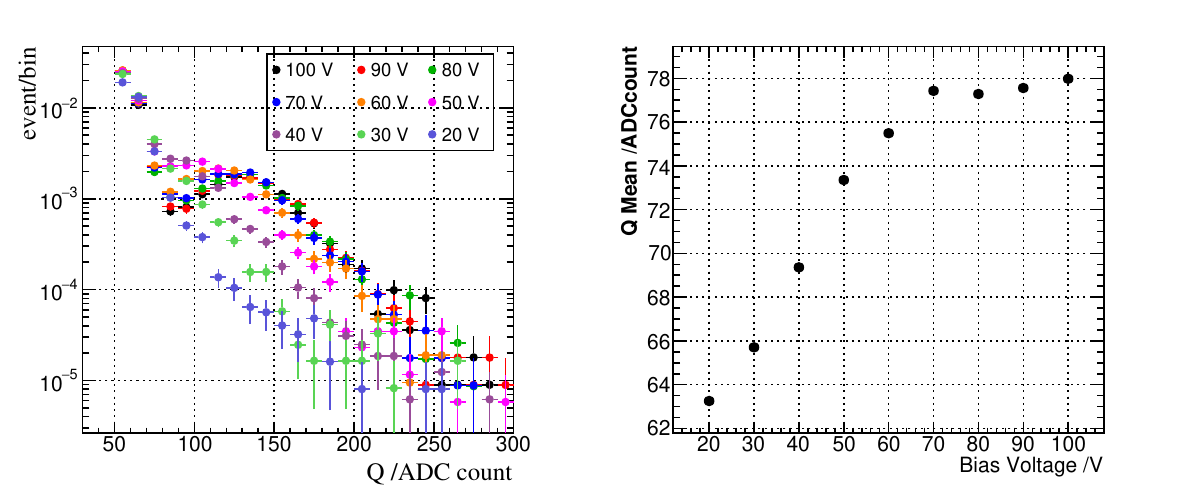}
    \caption{Top: charge distributions of MIP events at each bias voltage (left) and the mean of the charge distribution as a function of the bias voltage (right) for the X-SSD. Bottom: the same plots for the Y-SSD. \label{fig:qvcurve}}
\end{figure}

It has been reported that the Hamamatsu S13804 sensor reaches full depletion at approximately 60 and 70~V of bias voltage, and more than 100~V is required to reach full depletion if the sensor has damage~\cite{cite:ito}.
To ascertain the quality of the sensors with respect to full depletion voltage, the MIP charge value was measured as a function of applied bias voltage (Q-V response).
Figure~\ref{fig:qvcurve} shows the observed charge distributions at each bias voltage (left) and the mean of the charge distribution as a function of bias voltage (right) for the X-SSD (top panels).
The corresponding plots for the Y-SSD are also shown in the bottom panels of Fig.~\ref{fig:qvcurve}.
The results indicate that the charge reached saturation at 50 V in the X-SSD and 70 V in the Y-SSD, which correspond to the full depletion voltages for the respective SSDs.
These values are consistent with the results of the previously discussed confirmation process.

Furthermore, a comparison was made between the observed MIP charge and the expected value to ensure the reliability of the results. 
The observed charge, measured in units of fC, can be calculated as follows:
\begin{equation}
    Q = p_0 \times Q_\mathrm{ACD} + p_1 \left[ \mathrm{fC} \right],
    \label{eq:adc_calib}
\end{equation}
where $Q_\mathrm{ACD}$ is the ADC count of the obtained SSD signal, $Q$ is the corresponding charge.
Because no calibration of charge-to-ADC count correspondence ourselves, we used the parameters $p_0$ and $p_1$ measured in Kobe University with a different APV module instead.
The calibration was done using a test pulse generated by an RC circuit with a given capacitor and 50~$\mathrm{\Omega}$ resistor, resulting in $p_0 = 0.030 \pm 0.001$ and $p_1 = -0.853 \pm 0.569$~\cite{cite:nagasaki}.
As a result, the MIP charge values observed by each SSD ($Q^{\mathrm{X}}_{\mathrm{MIP}}$ and $Q^{\mathrm{Y}}_{\mathrm{MIP}}$) with 100~V bias voltage were 
\begin{equation}
    \begin{split}
        Q^{\mathrm{X}}_{\mathrm{MIP}} &= 3.0 \pm 0.6~\left[ \mathrm{fC} \right], \\
        Q^{\mathrm{Y}}_{\mathrm{MIP}} &= 3.0 \pm 0.6~\left[ \mathrm{fC} \right],
    \end{split}
\end{equation}
respectively.
While the mean values were 20--30\% less than the expected value reported in~\cite{cite:g-2ssd}, they were within a tolerance level when taking the estimated error size into account.

\subsection{Beam profile and efficiency measurement}
For a demonstration of the prototype detector with the SSD, a profile measurement of the ARTBL beam was performed.
The position at which the beam electron traversed was reconstructed as a strip with the maximum charge on each SSD.
Note that the minimum requirement for charge identification is more than 100 ADC counts to accurately identify the MIP activity.
Consequently, two-dimensional track position can be determined by the coincidence of both SSDs.

\begin{figure}[tb]
\centering
    \includegraphics[width=0.49\textwidth]{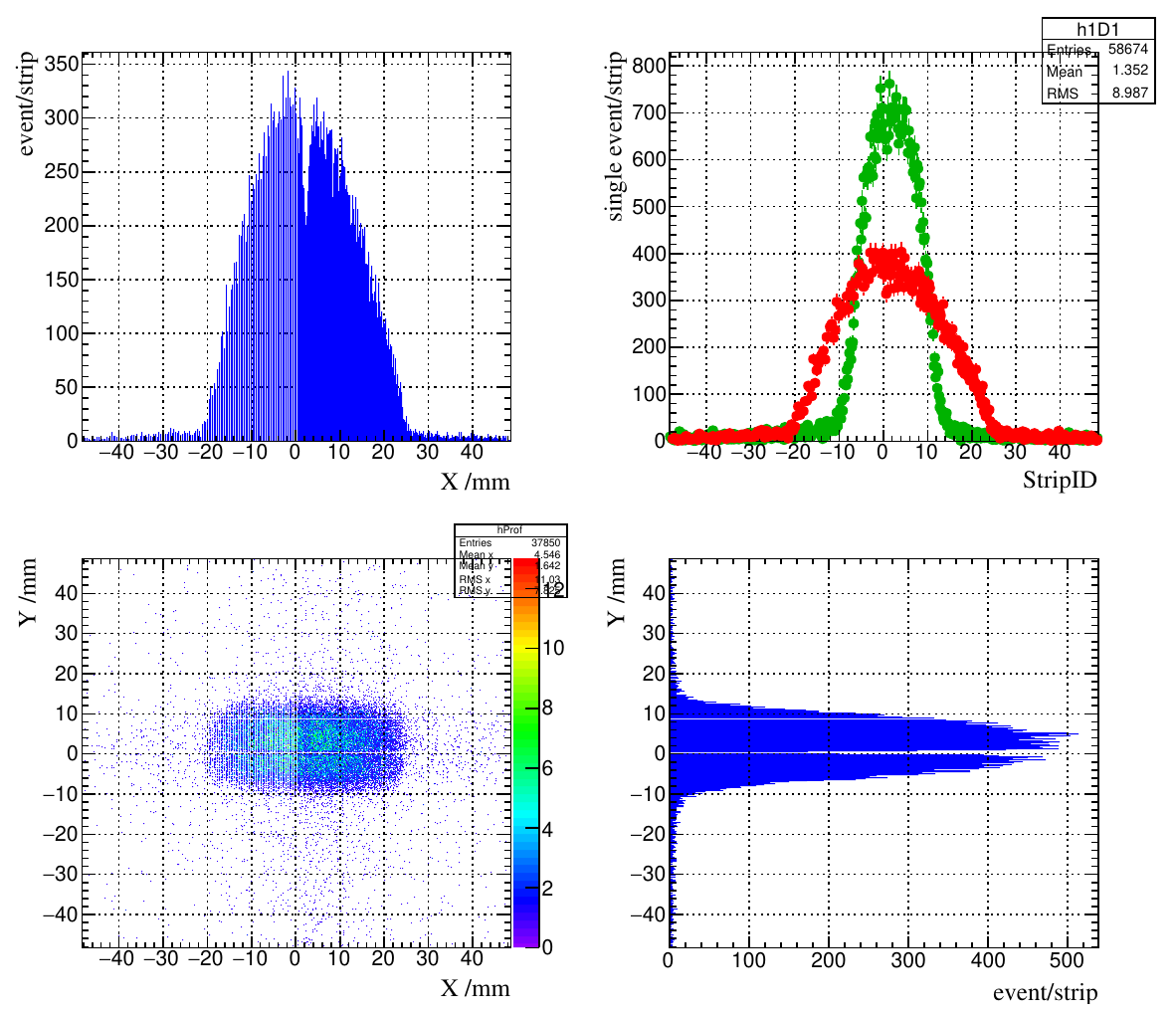}
    \includegraphics[width=0.49\textwidth]{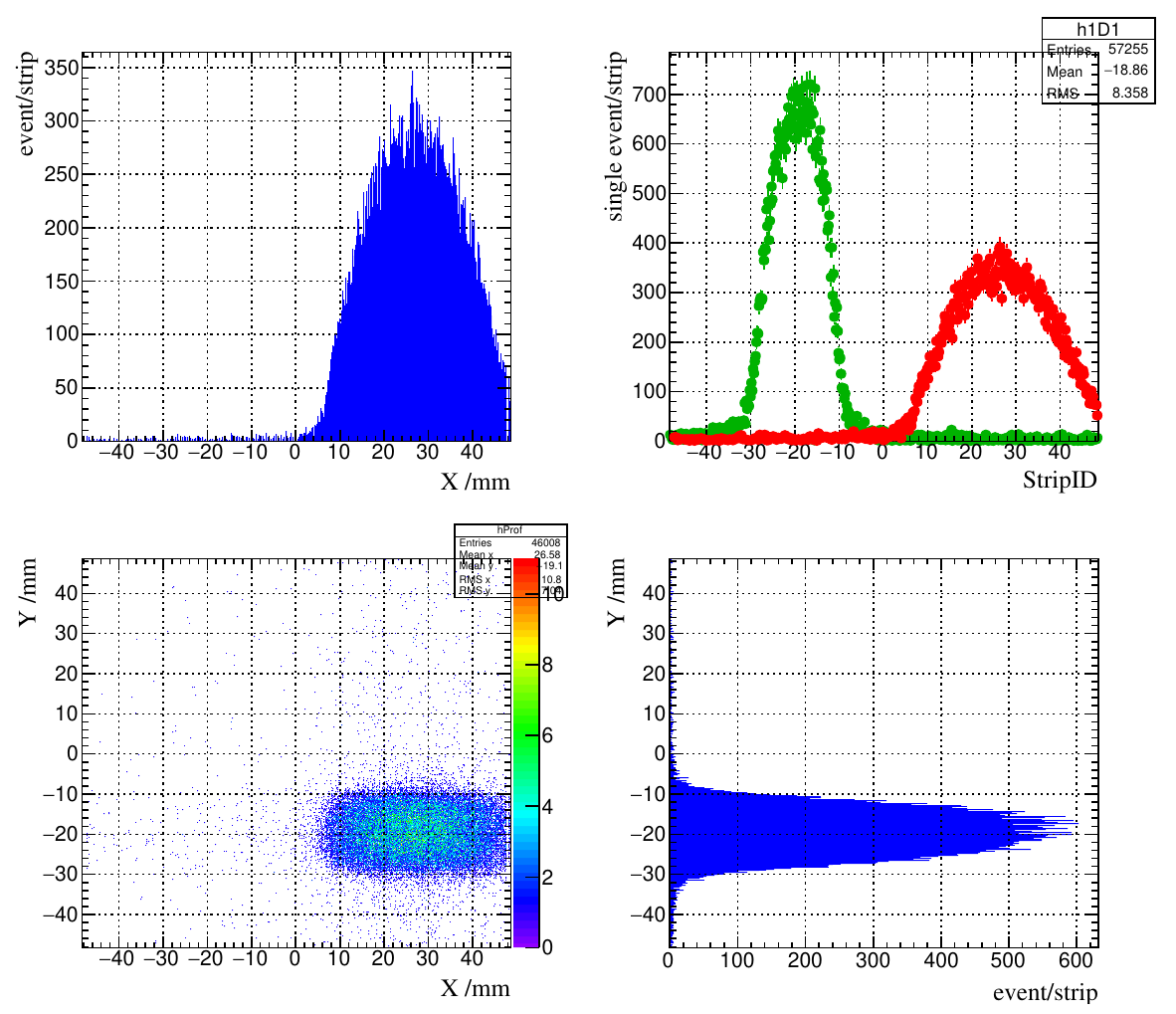}
    \caption{The results of profile measurement at (X, Y) = (0, 0 mm) (left) and (-25, +26 mm) (right) are displayed in four panels for each. 2D profile of the ARTBL beam measured by coincidence between X- and Y-SSD (bottom left), and the profile projected to X (top left) and Y axis (bottom right). The profile of X (red) and Y (green) was measured with the single SSD (top right). \label{fig:prof_artbl}}

\end{figure}

Figure~\ref{fig:prof_artbl} shows the results of the profile measurement at (X, Y) = (0, 0 mm) (left) and (-25, +26 mm) (right) in the coordinate system of the X-Y stage.
The figure consists of four panels for each measurement.
The bottom left plot of each side shows the 2D profile of the ARTBL beam measured by requiring coincidence between X- and Y-SSD, and the top left and bottom right plots are the profile projected to the X and Y axes, respectively.
The top right plot shows the event rate at a single strip corresponding to the profile of X (red) and Y (green) measured with the single SSD.
The measured beam profile's shape is consistent with the rectangular configuration of the lead collimator (60 mm width and 20 mm height), as expected. 
However, an effect of horizontal focusing due to a magnet parameter adjustment is visible.
The observation of an identical beam profile at the off-center location demonstrated the uniformity of the detection efficiency within the entire sensitive region of the silicon sensor.
The result of the beam profile measurement with the SSDs is consistent with the measurement using a multi-cell proportional chamber and a jet chamber in the ARTBL~\cite{cite:16aWA202-11}.
However, a region of reduced efficiency was identified along the line X = 0~mm and Y = 0~mm in the 2D profile, corresponding to the insensitive region of the sensor.
Note that this region was not observed in the profile at the stage position (X, Y) = (-25, +26 mm), as the beam spot was positioned sufficiently away from the insensitive gap.

The detection efficiency of MIP events at the SSDs was measured as a ratio of the number of MIP events in the SSD to that of the trigger, i.e.,
\begin{equation}
    \begin{split}
    \epsilon_{\mathrm{SSD}} &= \frac{R_{\mathrm{triple}}}{R_{\mathrm{trg}}} \\
                            &= \frac{R_{\mathrm{beam}} \times \epsilon_{\mathrm{PS1}}  \times \epsilon_{\mathrm{PS2}} \times \epsilon_{\mathrm{SSD}} }{R_{\mathrm{beam}} \times \epsilon_{\mathrm{PS1}} \times \epsilon_{\mathrm{PS2}}}.
    \end{split}
\end{equation}

Table~\ref{tab:si_eff} shows the measured detection efficiency of each SSD, as well as the coincidence of double SSDs, at the stage positions where the beam spot was centered and off-center relative to the detector. 
It should be noted that, in addition to the insensitive gap, the left half of the X-SSD strips was connected to only one APV hybrid due to a wire failure.
This resulted in a substantial decrease in the efficiency of the single X-SSD at the center relative to the efficiency at the off-center position.
In contrast to the case beam spot on the center, the beam spot avoided both the insensitive gap and the area affected by the wire failure at the stage position (-25 mm, +26 mm).
This resulted in more than 90\% of the efficiency at each single SSD and 89\% even in the coincidence between them.
The 4\% discrepancy in efficiencies is consistent with the finding that the edge of the beam is marginally outside the sensitive area of the X-SSD.
Consequently, it was determined that the prototype profile detector based on the SSD functioned effectively in the beam test.
However, a critical design update is imperative to address the loss due to the insensitive gap, a matter that must be addressed soon.

\begin{table}[bt]
    \centering
    \caption{Detection efficiencies of the MIP event with a single SSD and coincidence of double SSDs at the stage position (X, Y) = (0, 0 mm) and (-25, +26 mm), respectively. The statistical uncertainty of each value is less than 1\%. \label{tab:si_eff}}
    \begin{tabular}{cccc}\hline
        Stage Position (X, Y) /mm & X-SSD efficiency /\% & Y-SSD efficiency /\% & Coincidence efficiency /\% \\\hline
        (0,     0) & 75 & 90 & 68 \\
        (-25, +26) & 92 & 96 & 89 \\\hline
    \end{tabular}
\end{table}

\subsection{ToF measurement}
The time interval between the passage of beam particles through the scintillator detector was measured as the ToF.
The ToF resolution of the GAGG-based detector was evaluated as a touchstone for a PID performance in the low-E beamline. 

\begin{figure}
\centering
\includegraphics[width=1.00\textwidth]{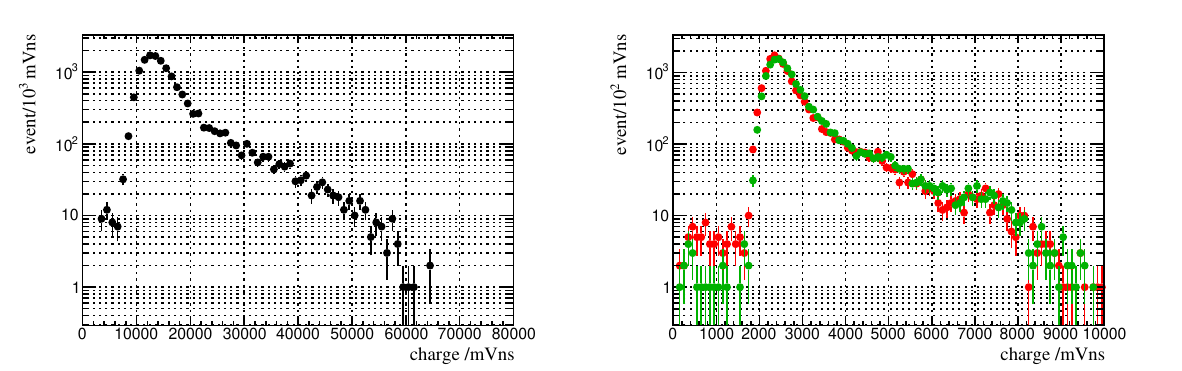}
\caption{Left: the observed charge distribution of the GAGG detector. Right: that of PS1 (red) and PS2 (green) in the unit of mV$\cdot$ns. Due to a limited dynamic range of the DRS4 evaluation board, the PMT gain for PS1 and PS2 was limited in order to keep linearity around the MIP events. \label{fig:charge_scint}}
\end{figure}

The signal waveforms of the scintillator detectors were digitized and recorded by the DRS4 evaluation board for each trigger.
Consequently, the charge and timing of each event were determined from the digitized waveform.
The baseline of each waveform was estimated as the mean of the first 50 samples.
The peak height of each waveform with respect to the baseline was defined as the pulse height of the waveform.
The observed charge was measured as an integral over 50~ns (100~ns) before and 250~ns (500~ns) after the peak timing for PS1 and PS2 (GAGG), taking the scintillation time constant into consideration.
The event timing was defined as the point at which the waveform reached one-third of the pulse height to address the timing-charge correlation.
Figure~\ref{fig:charge_scint} shows the distribution of observed charge for each detector exhibiting a Landau distribution, indicative of charged particle events.

\begin{figure}
\centering
\includegraphics[width=1.00\textwidth]{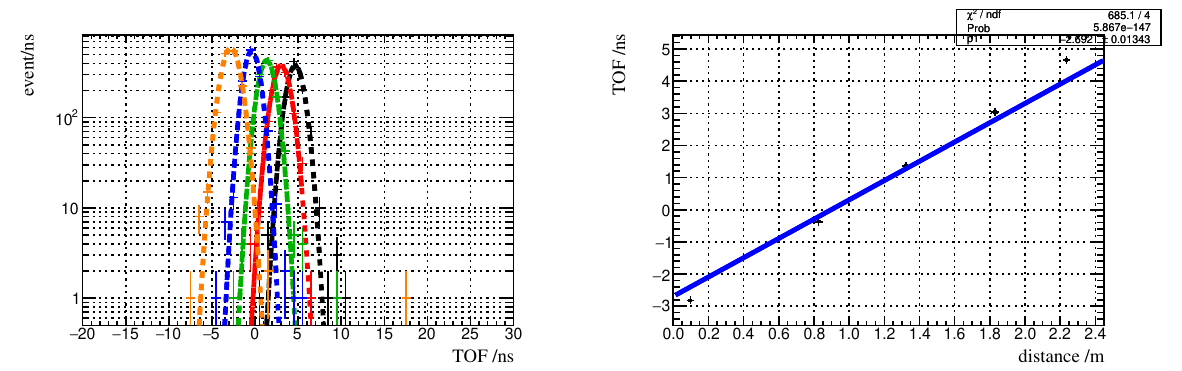}
\caption{Left: Distributions of ToF between GAGG and PS2 at 233.6 cm (black), 183.0 cm (red), 132.4 cm (green), 82.7 cm (blue), and 9.8 cm (orange), respectively. The results of the fit with a Gaussian are displayed as the dashed line. Right: Means of ToF as a function of the distance between GAGG and PS2. The blue line shows the result of a fit with a linear function. \label{fig:tof_gagg}}
\end{figure}

The ToF between the GAGG and PS2 was defined as the time interval between the event timings of the GAGG and PS2.
An event time spread was identified due to the limited number of observed photos around the peak of the Landau distribution.
Consequently, the events with more than 30,000~mV$\cdot$ns, corresponding to 100 photoelectrons in the rise part of the waveform, in GAGG, were selected for measurement of the ToF resolution instead of the entire MIP events.
Figure~\ref{fig:tof_gagg} shows the distribution of the observed ToF between GAGG and PS2 at each distance (left), along with the mean of the ToF distribution as a function of the distance between GAGG and PS2 detector (right).
While the measured ToF data demonstrate linearity with the detector distance, the fit with a linear function assuming the speed of light shows an inconsistency. 
However, it was determined that this factor does not impact the subsequent discussion on the ToF resolution estimation.

The ToF resolution of GAGG and PS2 was estimated as the standard deviation of the Gaussian distribution from the fit result. 
The mean of the ToF resolution was found to be approximately 900~ps, which exceeds the requisite resolution (500~ps) for the PID in the low-E beamline.
To achieve the planned PID performance with the GAGG-based ToF detector, it is imperative to enhance the light collection by an order of magnitude, as the evaluation demanded events with a high light yield instead of the entire MIP events.

\begin{figure}
\centering
\includegraphics[width=0.6\textwidth]{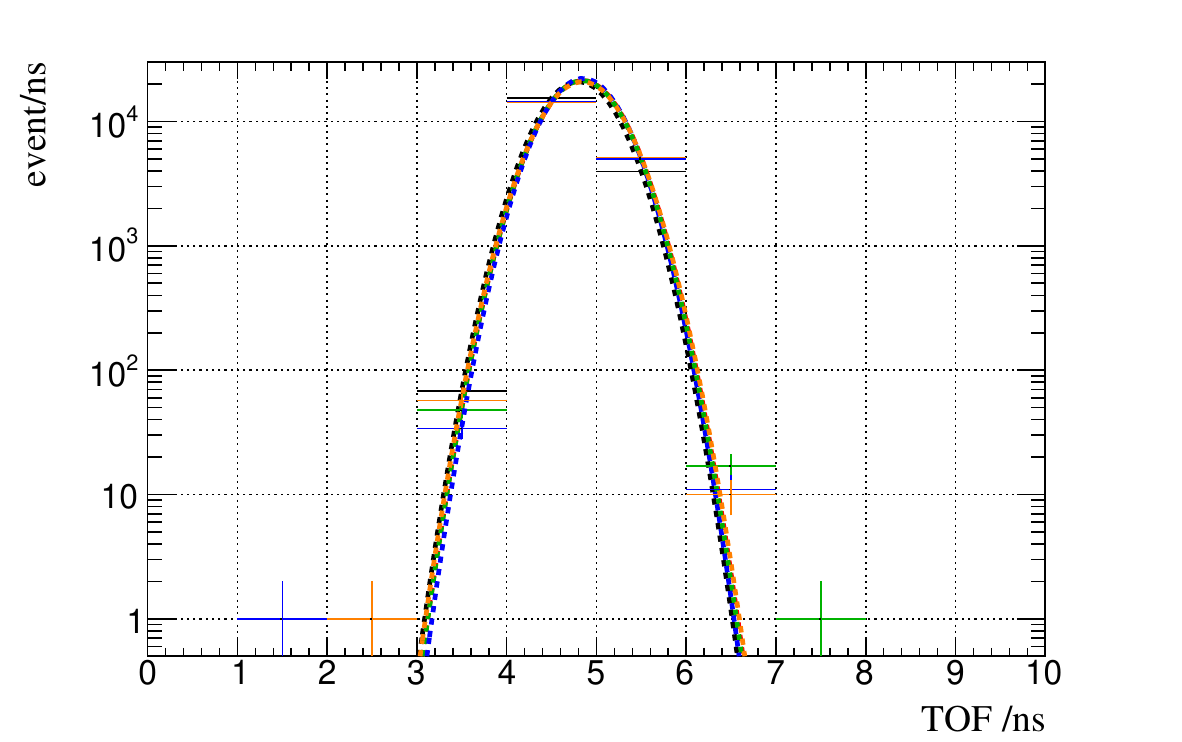}
\caption{Distributions of ToF between PS1 and PS2. The colors correspond to the same dataset with the configuration in Fig.~\ref{fig:tof_gagg} (left). The results of the fit with a Gaussian are displayed as the dashed line. \label{fig:tof_ps}}
\end{figure}

In addition to the GAGG-based detector, the ToF resolution of a conventional organic scintillator-based detector was evaluated using the ToF between PS1 and PS2.
Figure~\ref{fig:tof_ps} shows the distribution of the observed ToF between PS1 and PS2, with the color corresponding to the same dataset as shown in Fig.~\ref{fig:tof_gagg} (left).
It is noteworthy that the distance between PS1 and PS2 remains constant throughout the entire beam time, thereby ensuring the uniformity of the observed ToF values across different runs.
The ToF resolution was estimated using the same method as in the GAGG-based detector case, yielding an average of 390~ps over the runs, with a statistical error of approximately 1\%.
Consequently, it was determined that the organic scintillator-based detector fulfills the PID requirement for the low-E beamline.
In order to minimize beam loss at the detectors, 5~mm of plastic scintillator will be adopted for the ToF detector, resulting in a worse resolution than the one due to half the light yield.
However, it is possible to recover the resolution with photo-statistics and cancellation of timing uncertainty due to hit position by implementing a double-end readout with two PMTs.

\section{Conclusion}
Following the beam tests conducted at the KEK ARTBL in 2023, the performance of the prototype detectors for beam profile and time-of-flight was evaluated. 
The silicon strip detector demonstrated successful data acquisition with full strip readout using APV hybrids, and the measured two-dimensional beam profiles were found to be in agreement with those measured using a proportional chamber and a jet chamber. 
However, it was observed that the GAGG-based ToF detector exhibited suboptimal resolution in relation to the requirement for the PID in the momentum range of 2--13 GeV/c.
Conversely, the plastic scintillator-based ToF detector exhibited the necessary ToF resolution.
Consequently, plans are underway to engineer and evaluate a light collection enhanced ToF detector, characterized by a reduced scintillator thickness, with the objective of limiting further beam loss.

\section*{Acknowledgement}
We are grateful to KEK Instrumentation Technology Development Center for providing the test beam opportunities at the ARTBL and for their help with the beam tests.
We would like to thank for Dr. Tatsuya Chujo and Dr. Motoi Inaba for lending the electronics and hardware of the APV and SRS system.
This work was supported by Japan MEXT KAKENHI Grant Number 22H04943.

\end{document}